\documentclass[unnumsec,webpdf,modern,large]{mam-authoring-template}%

\usepackage{natbib}
\usepackage{amsmath,amssymb,amsfonts}
\usepackage{graphicx}
\usepackage{textcomp}
\usepackage{xcolor}
\usepackage{gensymb}
\usepackage{placeins}
\usepackage{siunitx}
\usepackage{upgreek}

\graphicspath{{Fig/}}

\begin{document}

\journaltitle{Microscopy and Microanalysis}
\DOI{DOI HERE}
\copyrightyear{2025}
\pubyear{2025}
\access{Advance Access Publication Date: Day Month Year}
\appnotes{Original Article}

\firstpage{1}

\title[A 4 megapixel HPCD for laboratory x-ray nanoCT]{Using a 4-megapixel hybrid photon counting detector for fast, lab-based nanoscale x-ray tomography}

\author[1,$\ast$]{Jordan Fonseca\ORCID{0000-0002-6325-3246}}
\author[2]{Zachary H. Levine\ORCID{0000-0002-9928-276X}}
\author[1]{Joseph W. Fowler\ORCID{0000-0002-8079-0895}}
\author[3]{Felix H. Kim\ORCID{0000-0002-1042-9500}}
\author[1]{Galen O'Neil\ORCID{0000-0003-3450-2263}}
\author[1]{Nathan J. Ortiz \ORCID{0009-0005-9384-0213}}
\author[4]{John Henry Scott\ORCID{0000-0002-3796-2110}}
\author[1]{Daniel S. Swetz\ORCID{0000-0002-8192-2175}}
\author[1,5]{Paul Szypryt\ORCID{0000-0002-0184-3419}}
\author[6]{Andras E. Vladar\ORCID{0000-0002-8823-6098}}
\author[1,$\ast$]{Nathan Nakamura\ORCID{0000-0001-8648-0524}}

\authormark{J. Fonseca et al.}

\address[1]{\orgdiv{Quantum Sensors Division}, \orgname{National Institute of Standards and Technology}, \orgaddress{\street{325 Broadway, Boulder}, \postcode{80305}, \state{CO}, \country{USA}}}
\address[2]{\orgdiv{Quantum Measurement Division}, \orgname{National Institute of Standards and Technology}, \orgaddress{\street{100 Bureau Dr}, \postcode{20899}, \state{MD}, \country{USA}}}
\address[3]{\orgdiv{Intelligent Systems Division}, \orgname{National Institute of Standards and Technology}, \orgaddress{\street{100 Bureau Dr}, \postcode{20899}, \state{MD}, \country{USA}}}
\address[4]{\orgdiv{Materials Measurement Science Division}, \orgname{National Institute of Standards and Technology}, \orgaddress{\street{100 Bureau Dr}, \postcode{20899}, \state{MD}, \country{USA}}}
\address[5]{\orgdiv{Department of Physics}, \orgname{University of Colorado Boulder}, \orgaddress{\street{2000 Colorado Ave}, \postcode{80309}, \state{CO}, \country{USA}}}
\address[6]{\orgdiv{Microsystems and Nanotechnology Division}, \orgname{National Institute of Standards and Technology}, \orgaddress{\street{100 Bureau Dr}, \postcode{20899}, \state{MD}, \country{USA}}}

\corresp[$\ast$]{Corresponding authors. 
\href{email:jordan.fonseca@nist.gov}{jordan.fonseca@nist.gov},
\href{email:nathan.nakamura@nist.gov}{nathan.nakamura@nist.gov}
}

\received{Date}{0}{Year}
\revised{Date}{0}{Year}
\accepted{Date}{0}{Year}

\abstract{Hybrid photon counting detectors (HPCDs) have unlocked new capabilities for x-ray-based measurements at synchrotrons around the world in the last 30 years. By leveraging independently optimized sensor and readout layers, they offer high quantum efficiency (\(> 80\) \%), ultra-low dark counts, sub-pixel point-spread function, and high count rates (\(> 10^{6}\) counts per pixel per second). Furthermore, their small pixel size and large active area endow them with excellent coverage and resolution for both real-space and reciprocal space imaging. Here, we demonstrate that HPCDs are also well-suited for laboratory-based nanoscale x-ray tomography (nano-xCT). We perform nano-xCT on an integrated circuit fabricated at the 130-nm node and produce a 3D reconstruction with 40 times more photons collected 20 times faster than in this group's previous work, for an overall speedup of 800\(\times\). We review the technical considerations of using an HPCD for tabletop tomography. We quantify our reconstruction image quality using well-established metrics, including the modulation transfer function (MTF), Fourier shell correlation (FSC), and contrast-to-noise (CNR), to validate our choice of experimental parameters that provide sufficient resolution and imaging speed. Using these metrics, we determine that even under current experimental conditions, 160 nm wiring features are reconstructed at 75-80 nm spatial resolution.   
\keywords{hybrid photon counting detector (HPCD), x-ray tomography (xCT), nanotomography, nano-xCT, integrated circuits (IC), critical dimensionality, scanning electron microscope (SEM), semiconductor metrology, failure analysis (FA)}
}

\maketitle

\section{Introduction}\label{intro}

Since their first observation in 1895, x-rays have revolutionized fields ranging from medical and industrial imaging to astronomy and materials characterization. Compared to visible light, however, x-rays are much more challenging to generate, manipulate, and detect. Because of this, in many cases, scientific and engineering breakthroughs in x-ray physics are ushered in by improvements to the hardware of x-ray generation and detection. 

An ideal x-ray detector offers high bandwidth and dynamic range, the ability to efficiently collect incident photons, and sufficient spatial resolution for imaging. For x-ray spectroscopy, energy resolution is an additional requirement. Since their first use at a synchrotron in 1999 for macromolecular crystallography (\cite{manolopoulos_x-ray_1999}), hybrid photon counting detectors (HPCDs) have offered a compelling alternative to other x-ray cameras such as scintillators or imaging plates, and they are now routinely used at synchrotrons around the world (\cite{miceli_application_2009, bronnimann_hybrid_2018}). In an HPCD, the photon is absorbed in a sensor layer with high quantum efficiency that converts incident x-rays to a cloud of electron-hole pairs.  A potential bias across the sensor layer then sweeps the charge carriers to collection electrodes at the back of the sensor layer, where each pixel is bump-bonded to an application specific integrated circuit (ASIC) that handles readout of the signal and counts incident photons. Because the processes of photon absorption and photon counting are, by definition, physically decoupled in HPCDs, each step can be independently optimized for specific conditions.

As has been pointed out in a review (\cite{forster_transforming_2019}), many of the traits that make HPCDs so useful for synchrotron experiments make them similarly useful for a variety of laboratory applications. HPCDs are attractive because of their high quantum efficiency, zero dark count, high framerate, high count rate, and adjustable energy thresholds. Their modular design with small pixels and large active areas makes them particularly useful for imaging applications, where the same attributes that improved the quality of macromolecular diffraction data at synchrotrons stand to make similar improvements to x-ray imaging in the lab. HPCDs have been used for medical imaging (\cite{giaccaglia_characterization_nodate, fardin_characterization_2023, taguchi_vision_2013}) and laboratory-based materials inspection such as x-ray computed tomography (xCT) (\cite{werny_elucidating_2022, solem_material_2021, muller_laboratory-based_2021, lutter_combining_2021}). 

Nanoscale xCT (nano-xCT) is a particularly interesting application of HPCDs since these fast, high-efficiency detectors enable significant improvements in imaging resolution and speed, both of which are in high demand in the semiconductor industry. By incorporating the information from different x-ray radiographs collected from many distinct perspectives, xCT generates a 3D reconstruction of an object (\cite{hounsfield_computerized_1973}). It is an established technique in medicine and industrial manufacturing to non-destructively identify the internal structure and uncover anomalies (\cite{withers_x-ray_2007, withers_x-ray_2021}). In medicine, spatial resolution on the order of 1 mm is typically sufficient, but for advanced manufacturing, spatial resolutions of microns or even tens of nanometers is highly desirable. In particular, the semiconductor industry now routinely produces 300 mm integrated circuit (IC) wafers patterned with billions of individual wiring elements which may each only be a few nanometers across. This application space calls for nondestructive critical dimensionality metrology capable of achieving both nanometer spatial resolution and rapid imaging and analysis speeds. As the desired spatial resolution of nano-xCT increases, so too does the required number of photons to produce a 3D reconstruction. Since laboratory-based xCT is typically photon-starved, state-of-the-art nano-xCT and ptychography measurements of ICs have taken place at synchrotrons (\cite{holler_x-ray_2014, holler_three-dimensional_2019}), where a high-flux source of coherent, monochromatic x-rays enables spatial resolutions down to 4 nm (\cite{aidukas_high-performance_2024}). 

While synchrotron-based nano-xCT measurements will always outperform lab-based techniques on the metrics of resolution and imaging speed, it is not practical to rely on them in many applications. For IC foundries, failure analysis requires measurement turn-around times on the order of one day, and thus having on-demand, in-house imaging at high speed and spatial resolution is essential. Currently, chip inspection that requires the highest spatial resolution possible is performed with electron microscopy techniques, such as focused ion beam scanning electron microscopy (FIB-SEM), where the region of interest is consumed and destroyed during the measurement process. There are already a number of commercial xCT tools available, but they typically offer spatial resolution in the \SI{300}{nm} to \SI{1}{\micro\metre} range (\cite{noauthor_tescan_nodate, noauthor_x-ray_nodate, noauthor_sigray_nodate, noauthor_zeiss_nodate}). There are ongoing academic efforts to develop in-house nano-xCT tools (\cite{nachtrab_nanoxct_2014, dreier_fast_2025}) for IC inspection without the need for the sample to be under vacuum, though these approaches achieve spatial resolutions in the \SI{300}{nm} to \SI{600}{nm} range. Resolution up to 177 nm have been reported, although these measurements were performed on slurry-phase catalysts, not on an IC chip (\cite{werny_elucidating_2022}). Lastly, there are alternate designs for scanning electron microscope (SEM)-based nano-xCT metrology on ICs that report spatial resolutions in the \SI{100}{nm} to \SI{150}{nm} range, albeit using more involved and destructive sample preparation (\cite{lutter_combining_2021}). Thus, while there are many promising ongoing research thrusts, the semiconductor industry currently has no great options for rapid, in-house, nondestructive imaging at spatial resolutions below \SI{500}{nm} (\cite{aryan_overview_2018}). Such a tool may offer unique utility for failure analysis of smaller, front-end-of-line features for which there are not currently good x-ray imaging options. 

In this work, we demonstrate the integration of an HPCD into a NIST-developed, in-house, SEM-based, nano-xCT tool (\cite{nakamura_nanoscale_2024}). We show that the HPCD's properties make it possible to collect data for a full 3D reconstruction of a \(\approx\)\SI{1200}{\micro\metre}\(^3\) region of an IC fabricated at the 130-nm node in just over 10 hours, with image analysis of our data indicating that comparable image quality could have been achieved in less than two hours. For comparison, data collection of the same sample region in the same tool with the previously used x-ray spectrometer took approximately 240 hours. We use well-established image quality metrics to confirm that our measurement resolution is sufficient to image the 160 nm features in the IC while maintaining a photon flux sufficient for fast measurement. We account for a number of considerations that make acquiring and processing tomography data with such a large-area detector unique, and we do so in a sufficiently generic way to be useful for a wide number of cone-beam xCT experimental apparatuses.

\section{Materials and Methods}\label{methods}
This work was performed using a custom SEM-based nano-xCT tool at NIST. The full system architecture with a different type of x-ray detector was first proposed in 2019 (\cite{lavely_integrated_2019}) and has since been implemented and thoroughly explained in \cite{nakamura_nanoscale_2024}. The reconstruction code used here, TomoScatt, is described in greater depth in \cite{levine_tabletop_2023} and \cite{levine_scatter_2019}. We provide a brief overview of the whole system here, but we direct the reader to those manuscripts for detailed methods of the SEM hardware or reconstruction algorithm and focus here on HPCD-specific considerations.

\begin{figure*}[!t]%
\centering\includegraphics[width=\textwidth]{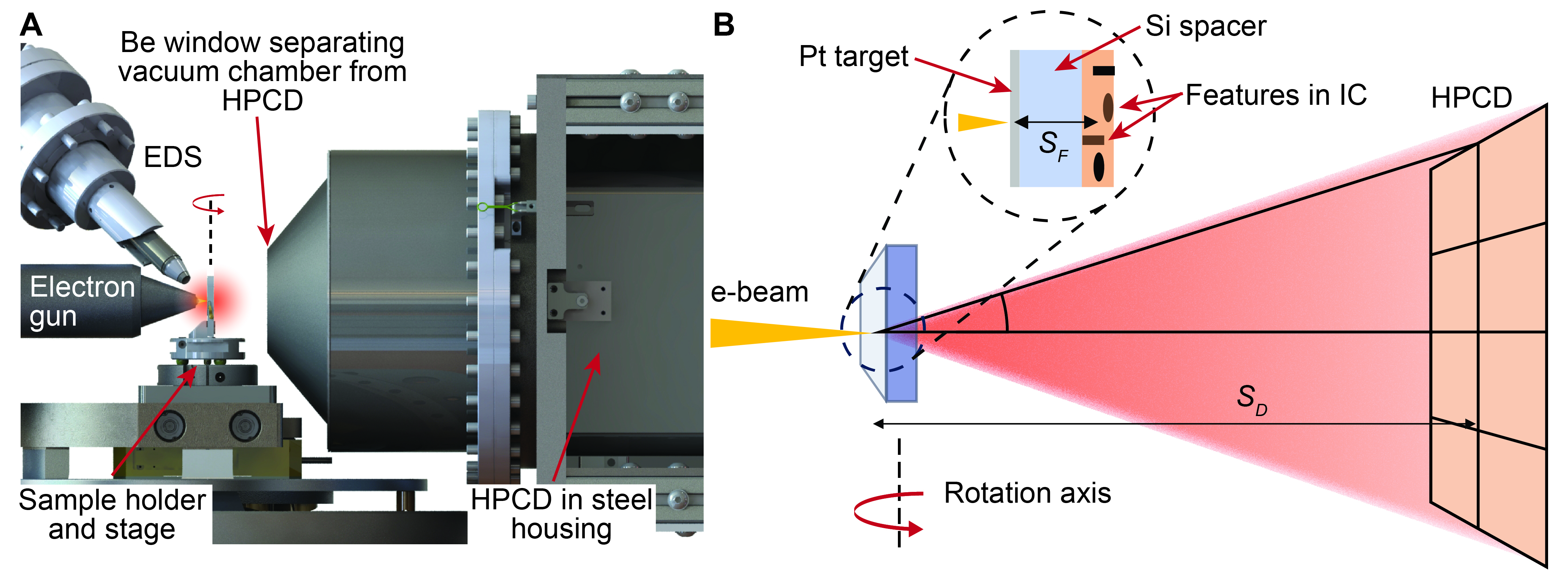}
\caption{Experimental schematic. (A) Computer assisted design (CAD) drawing of our SEM chamber showing the electron column, the stage and sample holder, the energy dispersive spectrometer (EDS), and the HPCD (placed outside the SEM vacuum chamber). (B) Cartoon side-view of the sample highlighting the interaction between the electron beam and the Pt target, the propagation of the generated x-ray spectrum through the IC sample, and the collection of the attenuated x-ray radiation by an HPCD. In cone-beam xCT, the source-detector-distance (\(S_D\)) and the source-feature-distance (\(S_F\)) determine the geometric magnification \(M_G = S_D/S_F\).} \label{fig_hardware}
\end{figure*}

\subsection{SEM Chamber and Sample Architecture}
The nano-xCT tool consists of a modified SEM vacuum chamber with a sample mounted on a fine positioning and rotation stage (Fig. \ref{fig_hardware}A). Incident electrons from the gun are converted to x-rays in a Pt target layer deposited directly onto the Si backing of the IC sample. For this measurement, the focused electron beam has a spot diameter of less than 100 nm, which leads to a comparably sized x-ray generation volume in the Pt target. The generated x-rays pass through the sample and exit the vacuum chamber through a Be window. These x-rays are collected by the HPCD detector, which is mated to the SEM chamber exterior with interlocks to prevent leakage of x-rays into the laboratory space.  

The sample, which is a Cu logic circuit fabricated at the 130-nm node, is thinned down until only six circuit layers, or \SI{3.32}{\micro\metre}, remain. The Si back of the sample is similarly thinned down to \SI{8.6}{\micro\metre} before a 100 nm layer of Pt is deposited onto this Si backing, as shown in the inset of Fig. \ref{fig_hardware}B The prepared circuit is then mounted to a glassy carbon substrate for structural support before the entire sample is attached to a graphite puck that can be mounted to the mechanical stage.

\subsection{HPCD}
The commercial HPCD used for this work is a DECTRIS Eiger2 R 4M. A number of key detector properties are tabulated in Table \ref{tab_hpcd}. When placed as close to the Be window as physically possible, the closest point on the detector active area is about \SI{256}{mm} from the point where the electron beam is focused onto the Pt target layer when the sample is at a 10 mm working distance from the end of the SEM column. Since this point is the source of the x-rays used for measurement, in the language of cone-beam CT, the system has a source-detector distance \(S_{D} = 256\) mm. The mean source-sample distance from the center of the Pt target to the center of the IC circuitry layer is \(S_F = \)\SI{10.26}{\micro\metre}. The system has a cone-beam geometric magnification of \(M_{G} = S_D/S_F\). At normal incidence, this gives a huge geometric magnification of \(M_G \approx 25~000\), corresponding to a projected HPCD pixel size of only \(3 \times 3\) nm. Because this is far smaller than any resolved features in the sample we have binned groups of \(16 \times 16\) pixels together in postprocessing to reduce the computational load of doing the reconstructions presented here.


\begin{table}[htbp]
\caption{HPCD Specifications. The final three parameters in the table depend on the sample rotation angle, with values at \(0\degree\) and \(22.5\degree\) shown here.}
\begin{center}
\begin{tabular}{|r|l|l|}
\hline
\textbf{Property [units]} & \multicolumn{2}{c|}{\textbf{HPCD}} \\
\hline
Active area [mm\(^2\)] & \multicolumn{2}{c|}{25~000} \\
Pixel array &  \multicolumn{2}{c|}{2068 \(\times\) 2162} \\
Pixel size [\SI{}{\micro\metre}\(^2\)] & \multicolumn{2}{c|}{75 \(\times\) 75} \\
Subtended angle(s) [\degree] & \multicolumn{2}{c|}{33.5, 35} \\
Solid angle [sr]& \multicolumn{2}{c|}{0.35} \\
\hline
& \textbf{HPCD (\(0\degree\))} & \textbf{HPCD (\(22.5\degree\))} \\
\hline
Geometric magnification & 25~000 & 23~100\\
Active area on sample [\SI{}{\micro\metre}\(^2\)] & 6.25 \(\times\) 6.53 & 7.32 \(\times\) 7.07\\
Pixel size on sample [nm\(^2\)]& 3 \(\times\) 3 & 3.51 \(\times\) 3.25\\
\hline
\end{tabular}
\label{tab_hpcd}
\end{center}
\end{table}

Prior measurements taken with this system used an x-ray spectrometer, which consisted of 240 pixels, each \mbox{\SI{350}{\micro\metre} \(\times\) \SI{350}{\micro\metre}}, contained within a disk 10 mm in diameter (\cite{pappas_tes_2019}). By contrast, the HPCD used for the measurements reported here contains over four million pixels and spans an active area of \SI{155.1}{mm} \(\times\) \SI{162.2}{mm}. Even held at a distance of 256 mm from the x-ray spot, this large, flat-panel detector introduces a number of geometric considerations that did not need to be addressed previously. There are two primary geometric effects that lead to a variation in apparent intensity across the detector plane of magnitude \((z/r)^3\) where \(z\) is the distance from the x-ray spot to the closest pixel on the detector and \(r\) is the distance from the x-ray spot to an arbitrary pixel. This intensity variation is due to the variation of solid angle coverage of each pixel on a flat array, and it arises from two distinct contributions. 

The first contribution is the reduction in solid angle coverage of each individual pixel due to distance. The solid angle of a point-like source subtended by a given pixel is \(\Omega(r) \propto 1/r^2\), so the reduction in solid angle coverage of an arbitrary pixel relative to the pixels closest to the source is \(\Omega(r)/\Omega(z) = (z/r)^2\). The second contribution is pixel obliquity, or how the amount of area of a given pixel that ``faces'' the x-ray spot decreases with increasing cone half-angle \(\phi\) away from the axis between the x-ray spot and the center of the detector according to \(\cos\phi = z/r\). These two effects multiplied together produce a \((z/r)^3\) geometric reduction in perceived intensity that must be taken into account to preserve the correct white-field photon statistics that are crucial for an accurate tomographic reconstruction.

In addition to these so-called ``geometric effects,'' that are independent of the x-ray detection mechanism, the probability of absorption within an HPCD pixel can also be affected by the path-length of the x-ray through the sensor layer, which depends on incidence angle. For a point-like x-ray source and a large flat-panel detector, this effect leads to energy-dependent changes to the absorption statistics across the HPCD face. This effect has been described in \cite{vaughan_characterization_2025}. Specifically, for a detector perpendicular to the beam source (which ours is), the correction factor takes the form 

\begin{equation}
k(E) = \frac{1 - \exp(-\mu(E) \rho t/\cos\phi)}{1 - \exp(-\mu(E) \rho t)}
\end{equation}
where \(\mu(E)\) and \(\rho\) are the energy-dependent x-ray mass attenuation coefficient and mass density of Si, respectively, \(t\) is the sensor layer thickness, and \(\phi\), defined above, is the angle between the beam axis and a given detector pixel (sometimes referred to as the scattering angle \(2\theta\) in diffraction experiments). These geometric corrections are summarized in Table \ref{tab_corrections}.

\begin{table}[htbp]
\caption{Geometric corrections for large flat-panel pixelated detectors}
\begin{center}
\begin{tabular}{|l|r|r|}
\hline
\textbf{Cause} & \textbf{Pixel on \(z\)-axis}& \textbf{Arbitrary pixel} \\
\hline
Distance effect &  \(1/z^2\) & \(1/r^2\)\\
\hline
Obliquity &  1 & \(z/r\) \\
\hline
Absorption prob. &  $1 - \exp(-\mu(E) \rho t)$  & $1 - \exp(-\mu(E) \rho t/\cos\phi)$ \\
\hline
\end{tabular}
\label{tab_corrections}
\end{center}
\end{table}

\begin{figure}[htbp]%
\centering\includegraphics[width=\columnwidth]{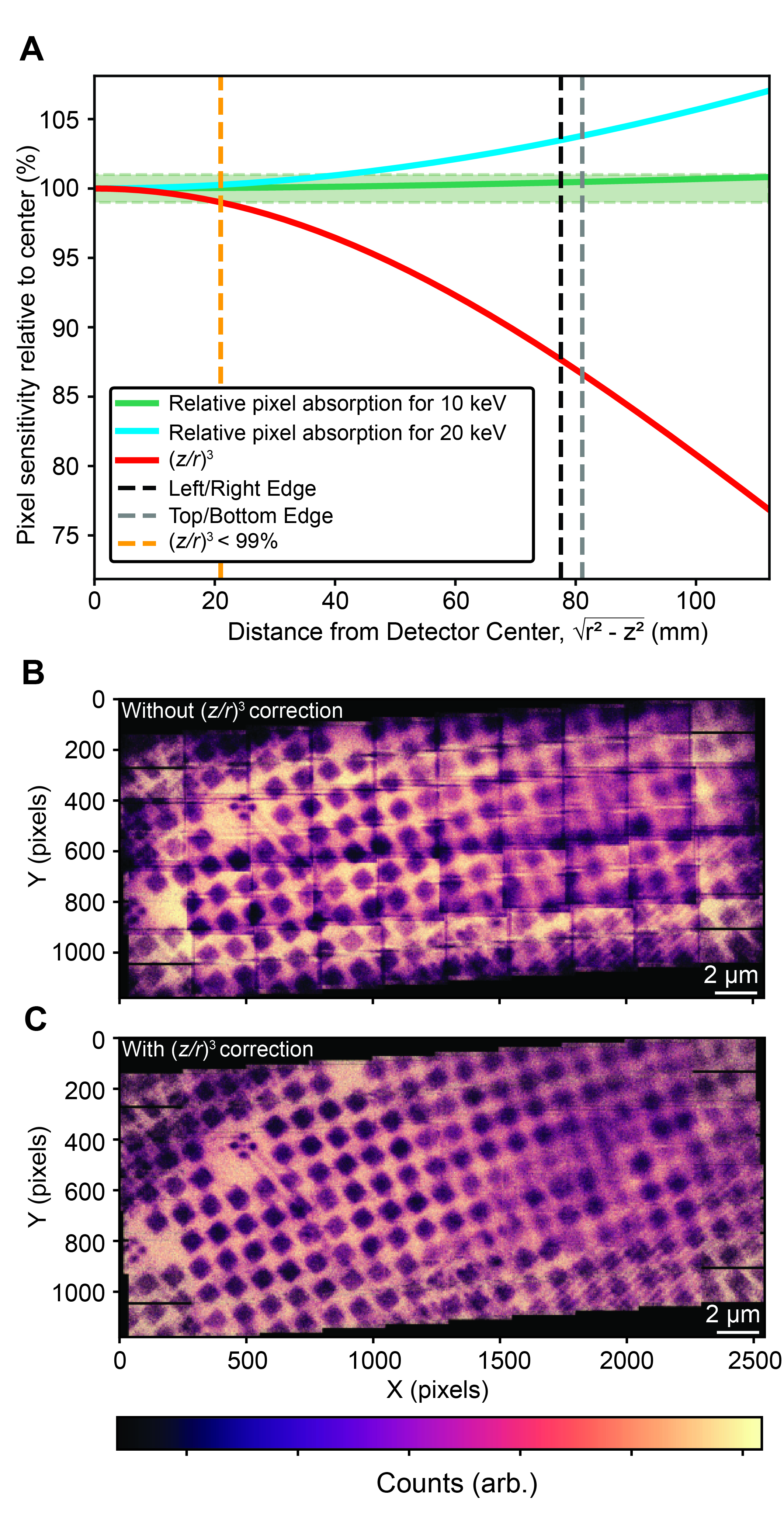}
\caption{Geometric considerations of a large flat-panel detector. (A) Plot of how pixel sensitivity changes as a function of distance on the detector surface from the pixel closest to the source. A purely geometric component \((z/r)^3\) is shown in red, while geometry- and energy-dependent change to the pixel absorption statistics for 10 keV and 20 keV are shown in green and light blue, respectively. Since the detector is rectangular, the plot shows the effect for pixels from the center all the way to a corner, with the edges of the detector marked by black and gray dashed lines. The green shaded area denotes a \(\pm 1 \%\) window that we consider to represent a negligible effect. This plot highlights the necessity of implementing a \((z/r)^3\) correction on our data while confirming that energy-dependent effects at 10 keV are negligible. Normal-incidence radiograph stitched together without (B) and with (C) this geometric correction implemented, highlighting the imaging artifacts that would be written onto the reconstructed dataset if the correction were omitted.}\label{fig_geom}
\end{figure}

Figure \ref{fig_geom}A shows a plot of these various geometric effects on our system. The red curve shows that the \((z/r)^3\) solid angle correction is dominant for our system, exhibiting a change in collection efficiency of over 20 \% between the center and the corners of the active area. By contrast, this plot highlights that for detectors with a diameter less than \(\approx20\) mm, the geometric \((z/r)^3\) correction can be omitted, explaining why this factor was not considered in prior work (\cite{nakamura_nanoscale_2024}). 

The green and light blue curves in Fig. \ref{fig_geom}A show the pixel-absorption correction factor at 10 keV and 20 keV, respectively. As the x-ray energy increases, the probability of a pixel absorbing the photon becomes more thickness-dependent and thereby more sensitive to the incidence angle. Over 80 \% of our photons lie between 4 keV and 10 keV, so we have omitted this correction factor from our data processing since it has a \(\leq 1 \%\) effect, as shown by the green curve in Fig. \ref{fig_geom}A. For similar tomography systems that use a larger detector, place the detector closer to the sample, or operate at higher x-ray energies, this energy- and angle-dependent variation in detector quantum efficiency can easily become a significant effect that needs to be corrected for during data acquisition and processing. 

To highlight the significance of correcting for the \((z/r)^3\) collection efficiency, we show example radiographs of the raw, 2D, x-ray transmission images, stitched together without (Fig. \ref{fig_geom}B) and with (Fig. \ref{fig_geom}C) this correction factor included. The same artifacts that cause visible intensity fluctuations and distortions in the uncorrected image here would cause significant artifacts in the reconstruction code if not corrected for in the input to TomoScatt.

\subsection{Limited Angle Tomography}

Our planar samples with a Pt target deposited on the back introduce constraints on how much the sample can be rotated in the course of collecting x-ray projections to use for tomography. Presently, to avoid collision between components in the SEM chamber, the circuit can only be rotated to \(\pm45\degree\). Even without this limitation, the Pt target would limit us to a rotation of somewhat less than \(\pm90\degree\). We address this challenge by performing limited-angle tomography. 

The data presented in this work were collected at only 7 discrete rotation angles spaced in increments of \(7.5\degree\) from \(-22.5\degree\) to \(22.5\degree\). At each discrete angle, the sample is mechanically translated by a piezo stage to let the beam dwell at a grid of points that tile the roughly \SI{25}{\micro\metre} \(\times\) \SI{10}{\micro\metre} scan area in steps of roughly \SI{1.5}{\micro\metre}, which is 1/4 of the projected detector width/height on the sample. The detector active area consists of pixel modules separated by small vertical and horizontal gaps where no photons are collected. The scan area is deliberately rotated in-plane by \(4\degree\) so that these gaps are not aligned for two adjacent images. See Supplementary Material Fig. 1 for a schematic of the scan plan.

One aspect of using a detector with large active area for cone-beam tomography that works in our favor is the range of angles subtended by the detector at any given sample rotation angle. For our measurement geometry, the HPCD subtends \(33.5\degree\) horizontally and \(35\degree\) vertically, as noted in Table \ref{tab_hpcd}. This means that as the sample is translated to let the SEM beam dwell at different positions in the ROI, we collect rays that pass through the sample through an effective \(33.5\degree\) arc. In this way, increased \textit{lateral} overlap between adjacent beam dwell spots translates to increased \textit{angular} coverage within the \(33.5\degree\) subtended arc. Hence, each discrete rotation angle includes a nearly continuous range of angles spanning \(\pm16.75\degree\) about that angle. In this way, even without rotating the sample at all, we can perform ``single-angle'' tomography. As the range of subtended angles \(\theta_s\) used for tomographic reconstruction increases, there is a corresponding increase in depth resolution that follows the simple trend \(1/\sin({\theta_s/2})\). So for single-angle tomography with \(\theta_s=33.5\degree\), the depth resolution will be 3.5 times worse than the lateral resolution. For the 7-angle data presented in Fig. \ref{fig_recon} and analyzed in Fig. \ref{fig_quality}, which have a subtended angle (rotation and detector combined) of \(\theta_s=78.5\degree\), the depth resolution should be only 1.6 times worse than the lateral resolution. See Supplementary Material Fig. 2 for an example of such a single-angle reconstruction, with slices that directly correspond to those for the 7-angle reconstruction shown in \mbox{Fig. \ref{fig_recon}}.

\subsection{Reconstruction Algorithm and Workflow}

All tomographic reconstructions in this work were performed with TomoScatt, a table-driven, physics-based reconstruction code developed at NIST (\cite{levine_tabletop_2023,levine_scatter_2019}).  The dataset shown in this work can be reconstructed in roughly one hour on a workstation using 8 processors coordinated by MPI code assuming a point-like x-ray source. The reconstructions presented here account for a distributed SEM spot; this extends the reconstruction time to \(\approx 24\) hours on a cluster, but is not strictly required to extract high-quality information from the reconstruction. In general, reconstructions are performed in a two-step process without reliance on any prior knowledge of the circuit. First, a robust Maximum Likelihood (ML) extremization is performed to arrive at a starting guess for the reconstruction that is likely close to the global ground truth. Then, using this guess, the algorithm is run Maximum A Posteriori (MAP) with a Bayesian prior to arrive at a high-fidelity, low-noise reconstruction of the sample. This two-iteration approach to the reconstruction prevents the MAP optimization from getting stuck at local extrema rather than the correct global maximum.

We also note our approach to the universal tomography challenge of inter-angle alignment. Accurate reconstruction of a 3D sample requires that the position of the sample be accurately tracked as the sample rotates. In nano-xCT, the alignment tracking requirements are stringent, since errors in displacement on the order of the voxel size (\(\approx 40 \) nm) will result in blurring. In this work, we perform 3D reconstructions of the sample at 7 discrete rotation angles. As mentioned above, since the detector subtends over \(30\degree\), each of the 7 rotation angles is itself a ``limited angle'' tomography measurement that can be used to get a coarse 3D reconstruction. Subsequently, individual slices taken from reconstructions at two adjacent measurement angles are aligned to each other, and a mean displacement between the two 3D reconstructions can be extracted and applied to guarantee that all the data in the full 3D reconstruction are as well-aligned as possible. By aligning the images based on partial 3D reconstructions, rather than the raw 2D x-ray radiographs, the overall alignment can be far better established since it is not subject to angular errors and inconsistent x-ray transmission associated with trying to align two different 2D projections.

\section{Results and Discussion}\label{results}

\subsection{Tomographic Reconstruction of Integrated Circuit}\label{reconstruction}

\begin{figure}[htbp]%
\centering\includegraphics[width=\columnwidth]{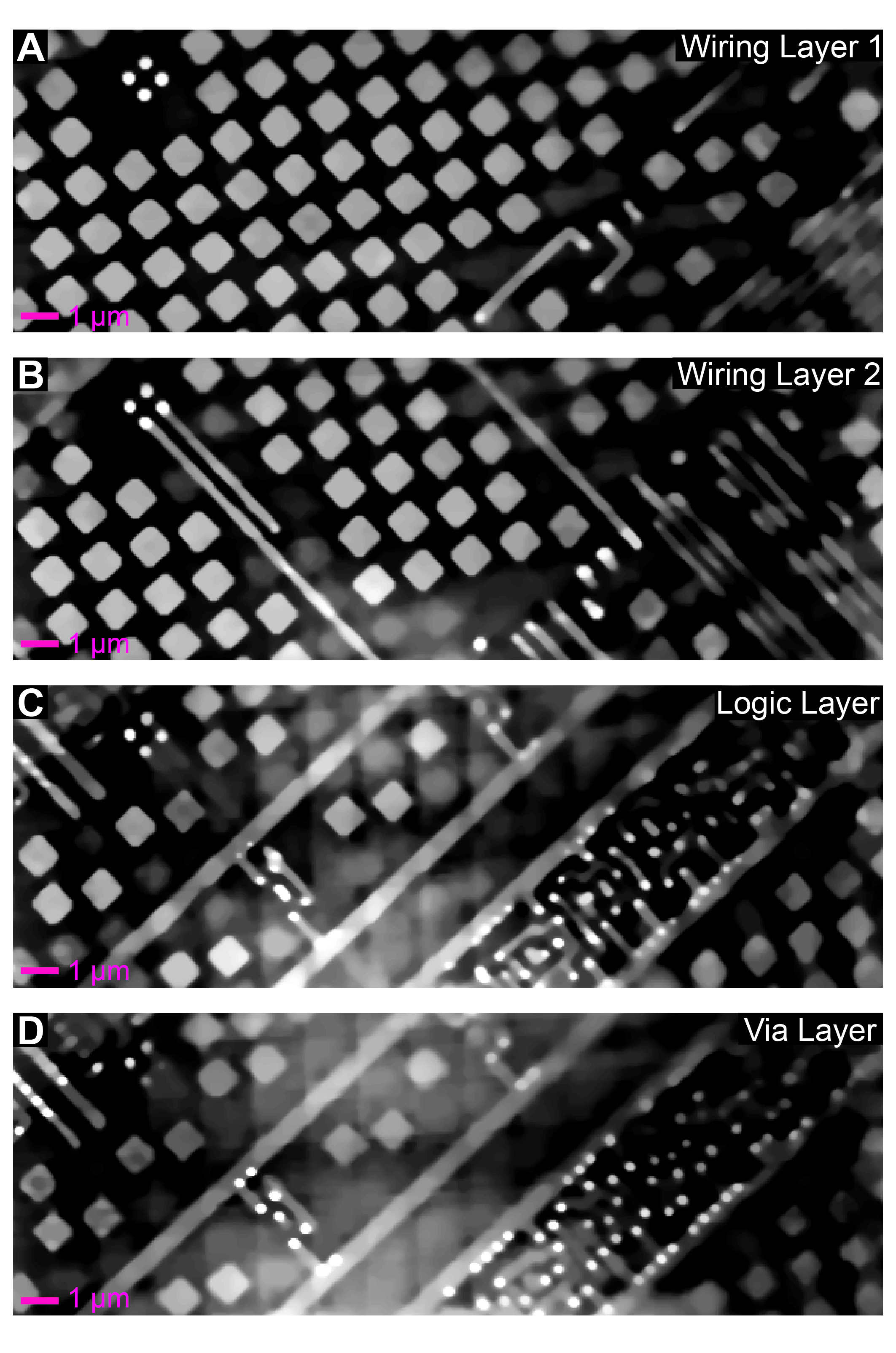}
\caption{Reconstruction results from a 7-angle dataset covering a \(78.5\degree\) arc collected with an HPCD. Two wiring layers (A,B), a digital logic layer (C), and a via layer (D) are all clearly distinguishable and well-resolved. Reconstruction voxels are 40 \(\times\) 40 \(\times\) 80 nm, so slices are 80 nm thick. From top to bottom, the slices depicted are separated by 640 nm, 560 nm, and 320 nm. The via layer contacts the Si wafer. Scale bar is \SI{1}{\micro\metre} in all images. Supplementary Material Fig. 4 shows a slice-by-slice comparison between the HPCD reconstruction and the spectrometer reconstruction (\cite{nakamura_nanoscale_2024}) for the four slices shown here.}\label{fig_recon}
\end{figure}

As a demonstration of the effective integration of an HPCD into our tomography tool, we image the same circuit reconstructed in \cite{nakamura_nanoscale_2024}. 
In Fig. \ref{fig_recon}, we show slices from four key layers of the circuit, including two different wiring layers, a digital logic layer, and a via layer that connects the logic to the Si gate. These data were collected in just over 10 hours of detector acquisition time, compared with the prior measurement's 210 hours. In that time, over 40 billion photons were detected, compared with the 1 billion photons used to reconstruct the circuit in prior work. Furthermore, we can downsample our dataset to show that we can reconstruct the circuit using less than two hours of data with no meaningful loss in spatial resolution (Supplementary Material Fig. 3). The slices shown in Fig. \ref{fig_recon} were reconstructed from a dataset of radiographs collected at 7 distinct rotation angles, covering a total limited-angle tomography arc of 78.5\(\degree\) (45\(\degree\) of measurement angles with a detector that subtends \mbox{33.5\(\degree\))}.

\subsection{Image Quality and System Resolution}\label{sec_quality}

In general, determining the resolution of an xCT system is a nontrivial task that depends on a large number of factors including experimental conditions as well as sample geometry and composition, not only equipment parameters. The x-ray spot size and SEM stability, depth-dependent geometric magnification, HPCD pixel size, positional stability during acquisition, hardware drift and vibrational stability, and reconstruction voxel size all contribute to the overall image resolution. The sample is rarely homogeneously reconstructed, particularly in limited-angle tomography, and some features or regions of the sample are almost guaranteed to provide higher resolution than others. Furthermore, the term ``resolution'' could refer to voxel size, HPCD pixel magnification, x-ray spot size, or any number of other imaging benchmarks.

In this work, we use a combination of the Modulation Transfer Function (MTF), Fourier Shell Correlation (FSC), and contrast-to-noise, as well as the benefit of having ground-truth information in the form of the design file, to comment on the quality of our images and the resolution capabilities of our \textit{experiment}. The resolution values presented can be used to get a sense of the capabilities of our instrument, but it is important to understand that the concept of ``resolution'' is a property of an instrument and sample together, not a property of the tool itself.  

\begin{figure*}[!t]%
\centering\includegraphics[width=\textwidth]{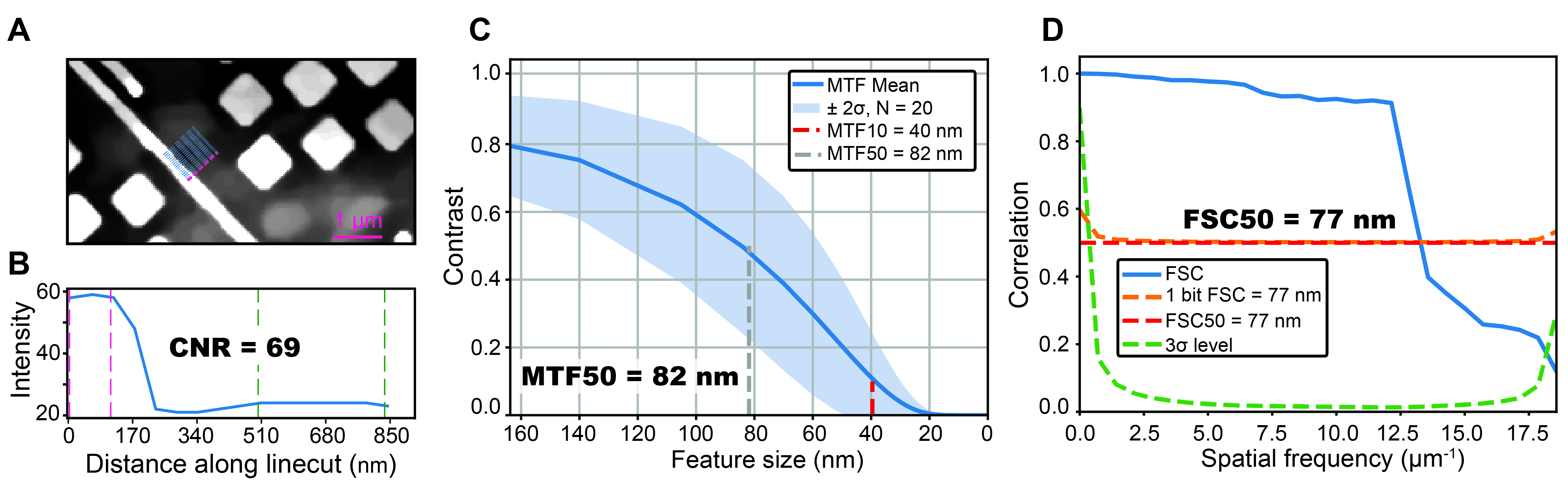}
\caption{Implementation of image quality metrics. (A) Cropped region of wiring layer 2 (Fig. \ref{results}B) showing a wire misaligned with the image pixel axis along which 20 lines have been drawn that will be used for MTF and CNR analysis. Scale bar is \SI{1}{\micro\metre}. (B) Example linecut from (A), showing change in reconstructed density between the wiring feature and the empty space adjacent to it. Pink and green vertical dashed lines denote two distinct regions on the linecut that are used for calculating CNR according to Equation \ref{eq_cnr}. (C) Plot of MTF contrast versus feature size in the sample. MTF curves are calculated for each of the 20 lines shown in (A), and the mean MTF curve is shown in dark blue, with the shaded region indicating a \(\pm 2\sigma\) confidence interval around the mean. MTF50 and MTF10 vertical lines highlight where the mean MTF contrast drops below 50 \% and 10 \%, respectively. We take the MTF50 threshold of 82 nm to be a conservative estimate of our system resolution to distinguish wiring from the background. (D) FSC curve of correlation versus spatial frequency (see Equation \ref{eq_fsc}), calculated from two completely independently reconstructed halves of a single dataset. The 1 bit per voxel and FSC50 thresholds are indicated by the orange and red dashed lines, respectively, with values 3\(\sigma\) above the noise floor indicated by the green dashed line. The FSC50/1 bit thresholds give a reconstruction-wide 3D resolution of 77 nm (the inverse of the spatial frequency threshold), consistent with our voxel size of \mbox{40 nm \(\times\) 40 nm \(\times\) 80 nm.}}\label{fig_quality}
\end{figure*}

In addition to making it possible to more quantitatively benchmark our instrument against alternative metrology systems and to gauge its usefulness in resolving a particular IC, we can use image quality metrics to confirm that our experiment is not resolution-limited in imaging the circuit at hand. This ability is valuable because while high spatial resolution is important, so too is performing the measurement as quickly as possible. Since the choice of SEM spot size presents a \textit{significant} tradeoff between photon flux and resolution, quantitative metrics that confirm that we are achieving sufficient resolution to image at the 130-nm node without compromising imaging speed is a critical development for our tool. A comprehensive overview of our reconstruction quality metrics is presented in Fig. \ref{fig_quality}.

By reporting the full frequency-dependent MTF and FSC curves as well as a variety of common cutoff thresholds for each metric, it is our hope to convey the general imaging capabilities of our instrument on this type of common IC sample without (incorrectly) suggesting that this capability can be boiled down to a single resolution number. See \cite{englisch_expanding_2023} for a good review of quantifying image quality in x-ray tomography.

Most simply, while signal-to-noise is a common image quality metric, in tomography, it is \textit{contrast} between foreground and background features within a given image that is meaningful. CNR is calculated according to 

\begin{equation}\label{eq_cnr}
    \text{CNR} = \frac{|\mu_1 - \mu_2|}{\sqrt{\sigma_1^2 + \sigma_2^2}}
\end{equation}
where \(\mu_1\) (\(\mu_2\)) is the mean pixel value in the foreground (background) and \(\sigma_1\) (\(\sigma_2\)) is the standard deviation in the foreground (background). Figure \ref{fig_quality}A shows the cropped portion of the IC slice shown in Fig. \ref{fig_recon}B with parallel linecuts drawn across the prominent wiring feature. The linecut extracted from the pink dotted line is shown in Fig. \ref{fig_quality}B, with the pink and green dashed lines denoting portions of the linecut that were used to determine the foreground and background mean and standard deviations for use in Equation \ref{eq_cnr}. While there are again many possible choices of threshold for what CNR is ``good,'' we note that the calculated CNR value of 69 for the wire presented is significantly above the widely used and probably somewhat conservative Rose Criterion (\cite{rose_sensitivity_1948}), which is an argument from both human perception and statistical inference that image features with \(\text{CNR}>5\) are clearly distinguishable from the background. CNR is a useful addition to the MTF for quantifying not only what spatial frequencies across a line are being transferred, but quite simply how ``visible'' a given feature is in the data.

Even though MTF analysis is necessarily performed on 2D images, it is the ISO standard for quantifying the sharpness of xCT reconstruction slices (\cite{noauthor_iso_nodate}) and has been used for the past 50 years in medical CT image analysis (\cite{schneiders_computer_1980,metz_transfer_1979, muller_laboratory-based_2021}). In this work, we apply the following process to experimentally determine the MTF curve for a given feature in a slice: (1) Take a linecut across a high-contrast edge (Fig. \ref{fig_quality}A,B). (2) Fit the linecut to a sigmoidal curve to obtain a noise-free edge spread function (ESF), as first proposed by \cite{nickoloff_simplified_1985}. (3) Differentiate the fitted curve to get a line spread function (LSF). (4) Take the modulus of the Fourier transform of the LSF to get an MTF curve. This process can be repeated for many different adjacent linecuts along an edge within a single slice (Fig. \ref{fig_quality}C), or across multiple slices to determine an average MTF curve and corresponding error bars for a feature within a slice or across several slices throughout the reconstruction. 

Fourier Shell Correlation is a standard technique in cryogenic electron microscopy (cryo-EM) of biological samples. It has been applied to xCT measurements, albeit not nearly as widely (\cite{englisch_expanding_2023}). FSC provides a normalized, spatial-frequency-dependent metric of similarity between two independently collected measurements of the same 3D sample. Formally, it is defined as 

\begin{equation}\label{eq_fsc}
    \text{FSC(k)} = \frac{\sum_{k_i,\Delta k} F_1 (k_i)\cdot F_2(k_i)^*}{\sqrt{\sum_{k_i,\Delta k} |F_1 (k_i)|^2 \cdot \sum_{k_i,\Delta k} |F_2 (k_i)|^2}}
\end{equation}
where \(F_1(k_i)\) and \(F_2(k_i)\) are the Fourier-transforms of two independently obtained 3D reconstructions at discretized spatial frequency shell \(k_i\). By taking two independent tomographic reconstructions collected under identical experimental conditions, FSC can be used to determine the spatial frequency at which point the two halves are no longer correlated above some threshold, which is a common metric of 3D, sample-wide spatial resolution. We perform FSC analysis on the same dataset presented in Fig. \ref{fig_recon} by splitting the acquired data into two datasets and reconstructing them independently under identical conditions before applying Equation \ref{eq_fsc}, with the results presented in Fig. \ref{fig_quality}D.

It is worth noting that it is not always possible to acquire two independent datasets, and there are proposed procedures for subdividing a single reconstruction volume in ways that make it possible to perform FSC based on a single measurement (\cite{verbeke_self_2024}). One contested aspect of using FSC for image resolution is choosing a cutoff frequency threshold to determine an experimental spatial resolution. While thresholds of \(0.143\) or \(0.5\) are common, there are statistical oversimplifications to relying on a constant threshold (\cite{van_heel_fourier_2005, van_heel_reassessing_2017}). As is clear in Fig. \ref{fig_quality}D, for our data, FSC50 (the resolution at which the FSC value drops below 0.5, or 50 \%) is equivalent to the 1-bit-per-voxel information content threshold presented in \cite{van_heel_fourier_2005} at the relevant crossover point and gives an FSC50 resolution of 77 nm. 

The significance of our quantitative analysis is twofold. First, the fact that our resolution is significantly better than the smallest features in our circuit gives us quantitative proof that we are imaging the actual features with fidelity. Second, these results emphasize that our current measurement does not push the limits of our experiment and suggests that even under the current experimental conditions, we could successfully reconstruct ICs with features down to approximately 75 nm. Furthermore, by reducing the SEM focal spot, we can expect to improve resolution further at the cost of reduced imaging speed. Our results, which are already on par with the state-of-the-art nano-xCT tools at a calculated resolution of \(<80\) nm with circuit features 160 nm wide clearly resolved, do not represent the limits of our technique. We anticipate that the next experiment with an upgraded SEM column and sample with smaller wiring features will deliver another factor of two in spatial resolution.

\section{Conclusion}

We have demonstrated the integration of a commercial HPCD into our SEM-based nano-xCT tool. Integration of the HPCD with our system required understanding how a large, flat-panel detector introduces imaging artifacts that must be corrected. The improvements in collection efficiency and spatial resolution enabled by the HPCD are especially well-suited to nano-xCT as it applies to modern semiconductor manufacturing, where the scale and complexity of modern ICs demand modern, in-house, nondestructive metrology techniques capable of addressing fault isolation and failure analysis questions with nanometer-scale resolution and single-day imaging speeds. The measurements presented here represent the state-of-the-art in laboratory-based nano-xCT metrology, clearly resolving 160 nm wiring features in mere hours of data acquisition and reconstruction time. 

The measurements taken with an HPCD were performed 20 times faster than in prior work while collecting 40 times more light. The reconstructions generated from these data show clear, easily resolvable circuit wiring for even the finest features in a chip fabricated at the 130-nm node. By applying standard 2D and 3D image quality metrics, namely CNR, MTF, and FSC, we determine a measurement resolution of \mbox{\SI{75}{nm} to \SI{80}{nm}}, indicating that our choice of experimental parameters was sufficient to guarantee that our reconstruction was not instrument-resolution-limited while preserving overall imaging speed. These quantitative tools make it possible not only to demonstrate that we can image at least six times faster with no loss in image quality, but also to contextualize our results against other measurements and tomography tools in the literature and to explore future tradeoffs between resolution and speed for any given application. 

\section{Competing interests}
ZHL declares Z. H. Levine, ``Efficient Method for Tomographic Reconstruction in the Presence of Fresnel Diffraction,''  U. S. Patent No. 12,367,563, Jul. 25, 2025. All other authors declare no competing interests.

\section{Author contributions statement}
NN, DSS, ZHL, PS, GO, and JWF conceived the experiment. NO designed and fabricated the HPCD x-ray housing. JF and NN ran the experiments and collected the data, with input from AV and JHS. JF, NN, and ZHL performed tomographic reconstructions and data analysis. All authors discussed the results. JF wrote the manuscript, with input from all authors.

\section{Acknowledgments}
The authors thank all team members who contributed to the development of and data collection using the proof-of-concept instrument, from both NIST and Sandia National Laboratories. Thanks to Amber Dagel, who led the fabrication of the original sample that was reused in this work. This work was performed with funding from the CHIPS Metrology Program, part of CHIPS for America, National Institute of Standards and Technology, U.S. Department of Commerce. Mention of commercial products does not imply endorsement by the authors or their institutions. The views and opinions of authors expressed herein do not necessarily state or reflect those of the United States Government or any agency thereof.

\bibliographystyle{plainnat}
\bibliography{reference}

\end{document}